\documentstyle[11pt,aaspp4]{article}

\singlespace
\tighten

\newcommand{\gtsim}{\ {\raise-0.5ex\hbox{$\buildrel>\over\sim$}}\ }
\newcommand{\ltsim}{\ {\raise-0.5ex\hbox{$\buildrel<\over\sim$}}\ }

\def\simlt{\lower.5ex\hbox{$\; \buildrel < \over \sim \;$}}
\def\simgt{\lower.5ex\hbox{$\; \buildrel > \over \sim \;$}}

\paperid{35654}
\received{7 Dec 1996}
\accepted{11 Feb 1997}
\revised{4 Mar 1997}
\journalid{}{}
\articleid{}{}

\begin{document}

Final version, approved for publication in the $Astrophysical~Journal$
\newline
\newline

\title{Measurement of an AGN Central Mass on Centiparsec Scales:
Results of Long-Term Optical Monitoring of Arp 102B}

\author{Jeffrey A. Newman, Michael Eracleous \altaffilmark{1}, Alexei
V.  Filippenko, and Jules P. Halpern \altaffilmark{2}}

\affil{Department of Astronomy, University of California, Berkeley, CA
94720 -- 3411 \\ e-mail: {\tt jnewman@astro.berkeley.edu,
mce@beast.berkeley.edu, alex@astro.berkeley.edu,
jules@jester.berkeley.edu}}

\vskip 24pt

\altaffiltext{1}{Hubble Fellow} 

\altaffiltext{2}{Permanent Address: Department of Astronomy, Columbia
University, 538 West 120th Street, New York, NY 10027, e-mail: {\tt
jules@carmen.phys.columbia.edu}}

\begin{abstract}

The optical spectrum of the broad-line radio galaxy Arp 102B has been
monitored for more than thirteen years to investigate the nature of
the source of its broad, double-peaked hydrogen Balmer emission lines.
The shape of the lines varied subtly; there was an interval during
which the variation in the ratio of the fluxes of the two peaks
appeared to be sinusoidal, with a period of 2.16 years and an
amplitude of about 16\% of the average value.  The variable part of
the broad H$\alpha$ line is well fit by a model in which a region of
excess emission (a quiescent ``hot spot'') within an accretion disk
(fitted to the non-varying portion of the double-peaked line)
completes at least two circular orbits and eventually fades.  Fits to
spectra from epochs when the hot spot is not present allow
determination of the disk inclination, while fits for epochs when it
is present provide a measurement of the radius of the hot spot's
orbit.  From these data and the period of variation, we find that the
mass within the hot spot's orbit is $2.2^{+0.2}_{-0.7} \times
10^{8} $~M$_{\sun}$, within the range of previous estimates of masses of
active galactic nuclei.  Because this mass is determined at a
relatively small distance from the central body, it is extremely
difficult to explain without assuming that a supermassive black hole
lies within Arp 102B.

Our collection of spectra allows us to apply several tests to models
of the source of the double peaks.  The ratio of H$\alpha$ to H$\beta$
flux at a given velocity displays no turning points or points of
inflection at the velocity associated with the blue peak in flux;
thus, this peak should not correspond to a turning point in physical
conditions.  This behavior is consistent with simple accretion disk
and, possibly, spiral shock models, but not with models which attribute the
double peaks to separate broad-line regions around a binary black hole
or to broad, subrelativistic jets.  The lack of systematic change in
the velocity of the blue peak over time provides a further constraint
on binary broad-line region models; this yields a lower limit on the
mass of such a binary black hole system of at least
$10^{10}$~M$_{\sun}$.  The variability properties of the double-peaked
emission lines in Arp 102B therefore continue to favor an accretion
disk origin over other models.

\end{abstract}


\section{Introduction}

Although there is a broad consensus that disk accretion onto a
supermassive black hole provides the tremendous power of active
galactic nuclei (AGNs), direct evidence for that process has remained
elusive.  The accretion disks expected from theoretical models are too
compact to resolve in even the nearest galaxies, and any optical
emission lines that would display the dynamical signature of the disk
are evidently too faint to be detected easily, at least in most cases.
There does, however, exist a class of AGNs (the ``double-peaked
emitters'') whose spectra exhibit hydrogen Balmer emission lines
having two broad peaks (one redshifted and one blueshifted) widely
separated from the systemic velocity of their host galaxy (Eracleous
\& Halpern 1994).  In many other cases, such as the H I spectra of
spiral galaxies and the optical spectra of cataclysmic variables
(Marsh 1988; Young \& Schneider 1980), this kind of line profile is
the spectroscopic ``signature'' of gas rotating in a disk.  One of the
first AGNs in this class to be so identified was Arp 102B, a
broad-line radio galaxy at a redshift of 0.02437 (Halpern \&
Filippenko 1988).  Models of either circular or elliptical
photoionized accretion disks typically give good fits to the Balmer
line profiles of this and other, similar objects.  Furthermore, an
accretion disk provides the most straightforward interpretation of the
absence of double-peaked components in high-ionization broad lines
(Halpern $et~al.$ 1996).

Alternative models that may produce double-peaked emission lines have
been proposed.  A binary black hole with broad line emission peaking
around each member (Gaskell 1983), a wide (in opening angle)
subrelativistic bipolar outflow (Zheng, Binette, \& Sulentic 1990), or
emission from spiral shock waves within a disk (Chakrabarti \& Wiita
1994) may all produce spectra with multiple peaks displaced from the
systemic velocity.

Changes in the emission-line profiles over time may allow
discrimination among these models.  For instance, in the binary
broad-line region (BLR) model, the two peaks should behave as a
double-lined spectroscopic binary; determination of the period and
velocity amplitude of this variation would constrain the combined mass
of the central objects.

If a disk origin for the emission lines were established, the patterns
and time scales of variability might provide information about the
structure and behavior of disks.  The line profiles of elliptical
disks will change their shape in a characteristic way as the disk
precesses (Eracleous $et~al.$ 1995), while other variations may
indicate the presence of inhomogeneities in the disk.  Simple spiral
shock models sometimes produce a third peak at lower velocity and
predict patterns of evolution of the line profile (Chakrabarti \&
Wiita 1993, 1994).  Detected reverberation might constrain all models,
since it could specify the length scales involved.  More detailed
reverberation observations could provide even stronger limits on
models of the structure of the BLR (Blandford \& McKee 1982; Stella
1990).

One feature of the spectrum of a circular accretion disk is that the
blue peak is brighter than the red one due to relativistic boosting
(in this paper, ``blue peak'' refers to the peak with smaller
wavelength).  Miller \& Peterson (1990) have asserted that, since one
Lick image dissector scanner (IDS; $cf.$ Robinson \& Wampler 1972)
spectrum of Arp 102B appeared to have a higher red peak than blue, the
accretion disk hypothesis is excluded.  However, as will be shown
below, a simple non-axisymmetric disk model can easily reproduce such
profiles.  We note that none of our spectra, including ones taken
roughly two months before and a year after theirs, contains such a
strong, rounded feature in the Balmer lines.  Furthermore, our spectra
with narrow lines removed appear to peak at a substantially different
wavelength from the peak found by Miller \& Peterson (an average of
about 6695 \AA\ versus roughly 6650 \AA\, respectively).  It is
possible that they in fact observed a particularly intense but very
short-lived ``hot spot'' of the sort described in \S 4.2, below.

Motivated by the possibility of testing the accretion disk and other
models, we have monitored the behavior of the optical emission-line
spectrum of Arp 102B from 1983 through the present.  Consecutive
observations were separated by intervals as short as a day and as long
as two years.  This paper presents the results of that campaign
through June 1996.  We model the double-peaked profile of H$\alpha$ as
emission from an accretion disk, in addition to evaluating the
applicability of the other models described above.  Our models of the
variation of the line profile produce a determination of the mass of
the central object directly from its gravitational effects on scales
of less than a centiparsec.  We describe the observations in $\S$ 2
and the measurements made using the data in $\S$ 3, apply accretion
disk models in $\S$ 4, examine the implications for other models in
$\S$ 5, and discuss the results in $\S$ 6.

\section {Observations} 

We have been observing the optical spectrum of Arp 102B since 1983
using several telescopes, as listed in Table 1.  Some of these spectra
have been published previously (Halpern \& Filippenko 1988, 1992;
Eracleous \& Halpern 1993).

The spectra of double-peaked emitters generally include appreciable
contributions from starlight as well as a substantial nonstellar
continuum component.  These were removed from the data by subtracting
a linear combination of normalized spectra of template galaxies (from
a previously collected library of spectra of E and S0 galaxies) and a
power law.  For the continuum around H$\alpha$, spectra of NGC 7332
and UGC 555 provided the best fit out of the library galaxies.  For
the continuum around H$\beta$, other template galaxies had to be used
as those spectra which fit best around H$\alpha$ did not extend to the
shorter wavelengths.  For both spectra around H$\alpha$ and around
H$\beta$, subtracting the best fitting combination of templates and
power-law resulted in a flat residual very close to zero flux.  Strong
stellar absorption features such as Na I D were largely removed, and
narrow He I $\lambda$5876 was typically visible in emission.

The fluxes of the starlight- and continuum- subtracted spectra were
normalized using the integrated flux in narrow [O III] $\lambda$5007
(for spectra around H$\beta$) or narrow [O I] $\lambda$6300 (for
spectra around H$\alpha$).  Since the narrow-line region of Arp 102B
is marginally resolved, variations in observing conditions and
extraction width among the observations yield different relative
amounts of the narrow- and broad- line fluxes included in the spectra;
thus, these normalizations are only approximate.  In this paper, no
results depend upon the absolute normalizations of the spectra, as we
interpret only the changes in the shape of the line.

\section {Data Analysis} 

\subsection {Relative Flux in the Two Peaks} 

It has been clear since the work of Halpern \& Filippenko (1988) that
the shape of the Balmer lines of Arp 102B varies significantly.  One
way to parameterize the evolution of the line profile is to measure
the relative fluxes of the blue and red peaks.  To estimate this, the
average flux between 6400 and 6520 \AA\ (in the rest frame of the AGN)
was used to measure the flux in the blue peak, while the average flux
in the combined ranges 6600 -- 6700 and 6760 -- 6780 \AA\ was
associated with the flux in the red peak.  These wavelength ranges
were selected to avoid all of the major narrow emission lines and the
range that the ``standard'' BLR (in H$\alpha$) would occupy ($cf.$
Halpern $et~al.$ 1996).  Significantly altering these ranges changed
the average flux by less than 1.5\% of the original value in each
case; thus, this method effectively measures changes in the ratio of
the $average$ fluxes in the two peaks.

To address the question of whether the maximum flux on the blue peak
exceeds that on the red peak in any spectrum, we would like to measure
the ratio of those quantities.  The [S II] lines prevent any precise
measurement of the flux at the top of the red peak, however.  To
estimate the ratio of the maximum fluxes of the two peaks, the average
flux ratio, measured as described above, was multiplied by a constant
determined using a previously published model fitted to one spectrum
of Arp 102B (Eracleous \& Halpern 1994) to approximate the actual
shape of the red peak of that spectrum (no physical significance of
that model is assumed at this stage).  Multiplication by this constant
does not necessarily yield the actual ratio of the peak fluxes for
each epoch.  However, the conclusions of this paper will not be
affected if the red-to-blue ratio (which, with this normalization, we
will refer to as $R$) is multiplied by any factor.

Figure 1 shows the history of the ratio $R$, determined using the
method described above.  At some epochs, $R$ approaches or may even
exceed unity, though the spectrum with the most extreme value of $R$ is
of lower quality than most (likely due to being the first following a
spectrograph replacement at Lick Observatory).  It is very difficult
to prove that the red peak in fact has the greater flux due to the
confusion of the narrow lines with that peak.  Nevertheless, it would
appear that the behavior of the H$\alpha$ profile of Arp 102B cannot
be completely explained by any $axisymmetric$ disk model.  It is
striking that there is an interval during which $R$ appears to vary
sinusoidally with an amplitude of about 16\% of the average value; the
variation begins in 1991 and seems to decrease in amplitude during
1995. Figure 1 presents a fit to those points designated with error
bars; the error bars shown are based upon the uncertainties in
determining the red and blue fluxes as described above.  The dashed line indicates the fitted curve
\begin {equation} 
R=R_0 + R_1 \cos {\left[{2 \pi \over P} (t-t_0)\right]}, 
\end {equation} 
where $R$ is the red-to-blue flux ratio, $P$ is the period of
variation, and $t$ is the Julian date, with $R_0=0.8637 \pm 0.0087,~
R_1=0.1371 \pm 0.0099, ~P=790 \pm 25$ days, and $t_0=2449634 \pm 23$
days. 
Compensating for cosmological time dilation, the period in the
rest frame of Arp 102B is $P / (1+z)$, or $771 \pm 24$ days.  This
sort of variation may not be unique to Arp 102B; Veilleux \& Zheng
(1991) have reported similar, albeit slower, variations in the ratio
of the fluxes of the two peaks of the H$\beta$ line of 3C 390.3.
However, no more than one cycle of that period was observed in 3C
390.3, and later data do not confirm it (Gilbert $et~al.$ 1997).
A similar variability pattern is also observed in 3C~332 (Eracleous
1997).  Possible implications of this variation of Arp 102B will be discussed in 
$\S$ 4.2.

\subsection {The Velocity of the Blue Peak}

The binary BLR model predicts that the two peaks should exchange their
positions relative to the rest wavelength with a period dependent upon
their combined mass, as in a double-lined spectroscopic binary.  The
expected period could be a few decades if the combined mass of the
binary is $\sim 10^8$~M$_{\sun}$.  We therefore measured the position
of the blue peak in our spectra of Arp~102B in search of the velocity
variations that this hypothesis predicts.  We have used Pogson's
method (see Gaskell 1996, and references therein) and fits to the peak
with Gaussian and quadratic functions for such measurements.  Pogson's
method yields results that are sensitive to substructure and skewness
in the peak, while Gaussian fits give a rough flux-weighted centroid
of the line.  The two methods are compared in detail and discussed
in a separate paper devoted to testing the binary black-hole
hypothesis in a number of double-peaked emitters (Eracleous $et~al.$
1997).

The results obtained using Pogson's method and the Gaussian fitting
method are given in Table~2.  In Figure~2$a$ we present these results
graphically by plotting the measured peak velocity (determined from
the wavelength relative to that of the narrow H$\alpha$ line using the
standard relativistic formula) as a function of time.  Uncertainties
in these measurements are typically of order 1~\AA, which translate
into uncertainties in the velocity of the peak of less than
100~km~s$^{-1}$.  However, systematic effects dominate the uncertainty
in the location of the peak.  In particular, the detailed shape of the
profile around the peak varies on time scales shorter than a year.  In
the binary BLR model, these variations could be attributed either to
reverberation of the individual BLRs in response to a varying ionizing
continuum or to changes in the distribution and/or velocity field of
the line emitting gas.  To bypass these uncertainties, we have
computed annual averages of the peak locations and have taken the
root-mean-square dispersion of the velocities for a given year to be
the uncertainty of their average.  Our own data were supplemented by a
measurement based on the spectrum of Stauffer, Schild, \& Keel (1983)
as reported in Halpern \& Filippenko (1988).

\subsection {The H$\alpha$ to H$\beta$ Flux Ratio} 

Although most of our spectra include only a wavelength range around
H$\alpha$, a smaller number include H$\beta$ as well.  The ratio of
the H$\alpha$ to H$\beta$ flux at a given velocity (hereafter $H$)
provides an indication of the physical conditions, particularly the
ionization parameter and to a lesser extent the temperature and the
density, of the emitting gas at that velocity.  It therefore may
constrain possible models for the source of the Balmer line emission.

Figure 3 shows $H$ for those years with enough spectra that included
H$\beta$ to allow significant reduction in noise through averaging.
The value of $H$ appears to vary monotonically across the entire blue
peak.  Though largely obscured by the presence of narrow emission
lines near either H$\alpha$ or H$\beta$, the behavior on the red peak
appears to also be consistent with a turning point at zero velocity
only.  Since there are no points of inflection or turning points in
$H$ at the velocity of the blue peak, it would seem that physical
conditions in the emitting material do not have a turning point at
that velocity.

This is precisely the behavior we would expect if an accretion disk
emits the double-peaked Balmer lines.  In that model, the locations of
the peaks correspond to the velocities with the greatest integrated
flux from the entire disk, not to a special region (such as the inner
or outer edge).  If the two peaks are produced by gas around each of
the black holes of a binary system, in contrast, one might expect $H$
to have a turning point where the H$\alpha$ flux reaches a peak, since
that presumably occurs at the velocity of one of the black holes;
admittedly, this scenario has not been modeled in detail.  One would
also expect a turning point in $H$ at the flux maximum in a model
where the two peaks correspond to broad jets, since the peak flux
should correspond to the most emissive parts of a jet.  The behavior
of $H$ within the spiral shocks model of Chakrabarti \& Wiita (1994)
is less clear.

\section {Application of Accretion Disk Models} 

\subsection {Circular and Elliptical Disks} 	 	

The variation of the two peaks that began in 1990 provides an
additional means of testing accretion disk models.  The difference
between an observed spectrum and a well-fitting model for disk-like
emission would be expected to include the narrow lines present and,
potentially, a broader component about H$\alpha$ resembling that of
typical BLRs.  The presence of a broad H$\alpha$ line from Arp 102B
originating in a region which is physically distinct from the source
of the double-peaked lines is an almost inescapable conclusion from
the ultraviolet line spectrum (Halpern $et~al.$ 1996).  We might also
expect our model fits to underestimate flux at the edges of the line
profile, since Gaussian broadening and a step-like decrease to zero of
the disk's emissivity as a function of radius are assumed to approximate
the combined effects of Compton scattering, turbulence, and physical
conditions that presumably vary more smoothly across the disk.  In
this paper, we demanded a closer fit of data to the models than in
existing published work.

In previous papers, models of circular, axisymmetric accretion disks
provided adequate fits to the presented spectra of Arp 102B (Chen,
Halpern, \& Filippenko 1989; Chen \& Halpern 1989).  However, as might
be expected from the discussion of the previous section, such models
cannot provide a good fit to the profile from late 1990 through 1995.
An axisymmetric disk model did provide an excellent fit to the
spectrum of 1990 July 17, the last spectrum of good quality before the
sinusoidal variation in $R$ began.  The parameters of that fit were:
power law index of the variation of emissivity with radius $q$ = 3.0;
line broadening given by convolution with a Gaussian with standard
deviation $b$ = 1050 km s$^-1$; disk inclination $i = 30.8^{\circ} \pm
1^{\circ}$, and inner and outer disk radii $\xi_i=305$ and $\xi_o=730$
gravitational radii (the gravitational radius $r_g={GM
/{c^{2}}}=1.4745 \times 10^{13} \, M_{8} \,$ cm, where $G$ is Newton's
gravitational constant, $M$ is the mass of the presumed central body,
$c$ is the speed of light, and $M_{8}$ is the mass of the central body
in units of $10^{8} \, M_{\sun}$).  These values were used as a basis for the
circular disk portion of the models described below.  Of these
parameters, only the inner and outer radii of the disk were allowed to
vary in later models ($q.v.$, below).

Following the inadequacy of the circular disk models, we attempted to
fit representative spectra that had extreme values $R$ with the
elliptical disk models of Eracleous $et~al.$ (1995).  These models
provided an excellent fit to those spectra with $R$ well below unity.
However, much poorer fits to those spectra with excess red flux were
achieved, and only then by allowing the eccentricity to vary from 0.1
to 0.59, the power law index $q$ to vary from 2.2 to 3, and $i$ to
vary from 28$^{\circ}$ to 31.5$^{\circ}$, all simultaneously.  Such
variations are more or less unrealistic.  For instance, Syer \& Clarke
(1992) have demonstrated that the eccentricity of an accretion disk
should change only slowly except in its innermost portions.
Inspection of their figures and the estimates of Eracleous $et~al.$
(1995) indicate that circularization by general relativistic
precession is effective only within about 100 gravitational radii,
while viscous circularization acts only over the very long viscous
time scale.  That, along with the fact that even with seven free
parameters only marginally satisfactory fits could be produced with $R
\approx 1$, would seem to indicate that the variations observed among the
spectra are not due to, for instance, a precessing elliptical disk.

 \subsection {Circular Accretion Disk with a Hot Spot} 

Eracleous $et~al.$ (1995) described the basic models of circular and
elliptical accretion disks adapted here; the original circular disk
models are described in Chen, Halpern, \& Filippenko (1989) and Chen
\& Halpern (1989).  The only significant addition to the previous
algorithm here is the inclusion of a region of the disk with excess
emissivity (hereafter called a ``hot spot'').  Line-emitting hot spots
at the intersection of a disk with an accretion stream have been
observed in the disks around cataclysmic variables (Marsh $et~al.$
1990), and a number of phenomena might produce them in an AGN
accretion disk.  In fact, a hot spot model for variation in the
emission-line profile of 3C 390.3 has been proposed by Zheng,
Veilleux, \& Grandi (1991).

We assume for our model the presence of a circular, rather than
elliptical, accretion disk to minimize the number of free parameters.
The hot spot is then implemented as an excess of emissivity within the
disk along a circular arc of infinitesimal extent in radius.  The
local viscous time scale, which is the time scale over which any sort
of instability may propagate radially, is larger than the period of an
orbit for radial extents of the instability above about $3 \times
10^{10} \alpha M_8^2 T_4$ cm at radii of interest, where $\alpha$ is
the traditional Shakura \& Sunyaev (1973) viscosity parameter and
$T_4$ is the temperature of the disk in units of $10^4$ K, applying
the definitions of time scales used below ($cf.$ equations 3 and 9).
Thus, even if the disturbance is not infinitesimal in radial extent,
its radial evolution should not be significant.  The excess emissivity
of the hot spot is assumed to vary as a Gaussian in azimuthal angle
along its arc away from the center of the hot spot.

Thus, the model requires four additional parameters (beyond those in a
circular disc model).  The first is $\xi$, the distance from the hot
spot to the central body in terms of the gravitational radius.  The
second parameter is $\theta$, the azimuthal angular position of the
hot spot's center measured within the disk.  When $\theta=270^{\circ}$, the hot spot has its greatest velocity towards the Earth.  The remaining parameters are
$\sigma$, the standard deviation of the Gaussian dependence of
emissivity on angle away from the hot spot's center, in degrees; and
$I$, a normalization factor representing the ratio of the hot spot's
maximum ($i.e.$, central) emissivity to that of the disc at the same
radius.  Thus, the rest frame luminosity of a hot spot 
relative to the disk luminosity is
\begin{equation}
 {L_{spot} \over L_{disk}} = {{\int_0^{\infty} \int_0^{2 \pi} I
 \xi^{-q} \,e^{-{(\theta - \theta_{spot})^2 \over 2 \sigma^2}} \delta
 (\xi -\xi_{spot}) \, \xi \, {d} {\theta} \, {d}{\xi}} \over
 {\int_{\xi_i}^{\xi_o} \int_0^{2 \pi} \xi^{-q} \xi \, {d}{\theta} \,
 {d}{\xi}}} = { \sigma I \over \sqrt{2 \pi} (\xi_o-\xi_i)} {\xi_i
 \xi_o \over \xi_{spot}^2},
\end{equation}
where $I$, $\theta_{spot}$, $\sigma$, and $\xi_{spot}$ are the hot
spot parameters described above (with $\sigma$ here given in radians),
and both $q=3$ and $\sigma << 2 \pi$ were assumed in evaluating the
integrals. 

In contrast to models with only a simple accretion disk, models of a
circular accretion disk with a hot spot provided an excellent fit to
all of the spectra during the time of variation in $R$.  We varied as
few parameters as possible in the course of the fits; excellent fits
were found for all spectra while varying only five parameters among
them.  The first of these are $\xi_i$ and $\xi_o$, the inner and outer
radii of the axisymmetric disk, which may be expected to vary
substantially because of changes in the ionization structure of the
disk (which determines the location of Balmer-line emission).  The
ionization state within the disk will respond to the amount of
ionizing flux very quickly (the recombination time scale in such a
disk is expected to be on the order of minutes, substantially less
than the light-crossing time).  That ionizing flux is presumably
determined by variability in the innermost parts of the disk, and thus
would also be expected to vary over much shorter periods than the time
for light to travel from the inner to the outer edge of the disk.
Thus, the inner and outer radii of the Balmer-emitting disk should not
be expected to vary in concert, but instead with relative (though
short) delays.

Similarly, we allowed the normalization of the hot spot flux, $I$, and
the standard deviation of its flux decline in angle, $\sigma$, to
vary.  Neither is determined by the fitting process to better than
perhaps 30\%.  We note that $\sigma$ varies over a substantial range
among all the models, and the hot spot intensity parameter $I$
decreases substantially after 1994.  Finally, $\theta$, the azimuthal
angle of the hot spot along its orbit about the disk, has to vary over
360$^{\circ}$ to be able to complete multiple cycles of
enhancing at some times the red peak and at others the blue.  Table 3
presents the results of these fits.  Note that in all cases, the hot
spot represented a perturbation of only a few percent in the total
Balmer line flux of the disk.  Due to the ambiguity of the helicity of
the hot spot's orbit from what are effectively radial velocity
measurements, there are two possible values of $\theta$ for each model
fit; only one is listed in the table.  Figures 4 and 5 display
representative model fits. Even in 1995, when the variation in $R$ is
no longer sinusoidal, the spectra are better fitted by a model with
some hot spot component (though with a lower normalization than
before) than by an axisymmetric disk.  Thus, it appears that the hot
spot gradually decreases in strength.

In all model fits plotted in this paper the hot spot was kept at a
radius $\xi$ of 455.  This radius was chosen based upon fits to data
in the interval 1991 June 17 -- 20, when the azimuth of the hot spot was near
270$^{\circ}$, so that its wavelength was shortest (and thus furthest
from the interference of narrow lines).  In fact, a good fit to that
spectrum was possible for models with $\xi$ from 355 to 485, but
beyond those bounds the fits were noticeably worse.  

That models with a single hot spot radius fit all of the spectra is
not in and of itself convincing; it is conceivable that some other
phenomenon might produce similar behavior.  However, if we adopt the
hot spot model, we know from Kepler's laws how a body must travel in a
symmetric potential along a circular orbit; in particular, it must
move with constant angular velocity in such a case.  Figure 6 plots
the possible value of $\theta$ nearer the line corresponding to the
phase of the sinusoidal variation in $R$ (taking the solution with
positive angular velocity) for each of the modeled spectra versus the
date of observation of that spectrum.  The depicted error bars of $\pm
30^{\circ}$ are a crude estimate based upon the modeling.  

It might be contended that the line fits so well so often simply by
virtue of having two possible points to fit at any given date (note
that only the possible $\theta$ which better fits the phase curve of
positive angular velocity is plotted).  However, if points are
distributed randomly, there is only a 1 in 3 chance that the line will
be within 30 degrees of either point for a given epoch.  Clearly, more
than the expected 7 points (namely, 14) are within their error bars of
the line; the probability of this many or more points falling so close
by chance is only 0.18\%.  We also note that residuals about the fit
line do not appear to occur systematically for a given phase at
different epochs; thus, the motion seems strongly consistent with
constant angular velocity.

We have determined the period of what seems to be an orbital motion
through the evolution of $R$ (or, alternatively, through the evolution
of $\theta$ in the models).  Indeed, the only proposed sources of
variability in an AGN that would cause simple sinusoidal variation in
$R$ with little apparent decay (in amplitude or frequency) for nearly
two complete cycles are orbital in nature.  We have also determined,
through our models, the radius $\xi$ (in terms of the gravitational
radius, $r_g$, which is proportional to the mass) associated with that
motion.  There are no ambiguities in the orbital inclination so long
as the ``hot spot'' is presumed to lie within the disk, which seems
very likely.  These data are, then, sufficient to find the mass within
the orbit of the hot spot (assuming a spherically symmetric mass
distribution) by the complete form of Kepler's third law for a
circular orbit:
\begin{equation} 
P=2 \pi \sqrt{r^{3}\over{GM}}={{2 \pi G M \xi^{3/2}} \over{c^{3}}}, 
\end{equation} 
where $P$ is the period of the orbit and $r=\xi r_{g}$ is its radius.

Using the above values for the period and radius of the hot spot,
we find $M=2.21^{+0.23}_{-0.73} \times 10^{8} $M$_{\sun}$,
within the range of previous estimates of masses of AGNs. For this
mass and $\xi =455$, the physical radius of the hot spot's orbit is
$4.80 \times 10^{-3}$ pc.  This corresponds to an average density for
a sphere with the same radius as the orbit of $2.38\times 10^{14}
$M$_{\sun} $ pc$^{-3}$.  Such a density is extremely difficult to
accomplish without the presence of a supermassive black hole.

\section {Evaluation of Alternative Models} 

\subsection {Binary Black Holes} 

We may expect the orbit of any binary black hole system to be
circular, as it should have evolved to its present state through
dynamical friction.  If we adopt an analogy between a binary black
hole system and a stellar, double-lined spectroscopic binary, the
velocities of the two peaks should vary sinusoidally.  Accordingly, we
have fitted the velocity variations of the blue peak with a curve of
the form
\begin{equation} 
v_{\rm obs}(t)= v\;\sin i \; \cos\left[{2\pi\over P_{\rm orb}}\left(t-t_{0}\right)\right]\; , 
\end {equation}
where $v_{\rm obs}(t)$ is the observed radial velocity of the blue
peak as a function of time, $v$ is the orbital velocity of the
corresponding black hole, $i$ is the inclination of the orbital plane
of the binary, $t_0$ determines the phase of the orbit, and $P_{\rm
orb}$ is the orbital period. The fitting algorithm used, which is
described by Eracleous $et~al.$ (1997), determines the confidence
intervals of the model parameters in addition to the best fit.

In Figure~2$b$ we show the variation of the annually averaged velocity
of the blue peak of the H$\alpha$ line of Arp~102B between 1982 and
1996 using measurements made with the Gaussian fitting method.  The
dashed line in this figure is the best fitting sinusoidal velocity
curve, which corresponds to a period of 390 years and an amplitude of
5200~km~s$^{-1}$.  The fit is quite poor, and the scatter of the data
points about the best-fitting curve does not seem random.  We find
that a sinusoidal modulation of the velocity of the blue peak must
have a period greater than 114 years to be even marginally consistent 
with the data. If the data from 1990 and 1991, when the centroid of
the blue peak seems to be varying rapidly, are excluded, the lower
limit to the period (and hence the mass) increases by a factor of 2.

The above observational constraints place a lower limit on the total
(combined) mass of the two black holes required by the binary BLR
model.  We note that the blue peak has a smaller velocity displacement
from the rest wavelength of the line than the red peak, which
associates it with the more massive of the two members of the
hypothesized binary.  Under this condition we can customize Kepler's
third law to produce an expression for the lower limit on the mass of
the binary ($cf.$ Eracleous $et~al.$ 1997),
\begin {equation} 
M > 4.7\times 10^8\; (1+Q)^3 \;
 \left({P_{\rm orb}\over 100\; {\rm yr}}\right)\;
\left({v\sin i\over 5000\; {\rm km\; s^{-1}}}\right)^3 \; {\rm
M_{\sun}},
\end {equation} 
where $Q$ is the mass ratio of the binary, estimated to be about 1.5
from the relative velocities of the two peaks.

The observational constraints derived above using all of the data
yield a lower limit to the total mass of $10^{10}$~M$_{\sun}$.
This is the lowest limit that the data will admit, and it is not
reduced if measurements made with Pogson's method are used.  
This bound is rather restrictive, since it is a
significant fraction of the mass of an entire galaxy, making the
binary black hole hypothesis unlikely for Arp~102B.  We stress that we
are only able to constrain the specific scenario in which each of the
two peaks of the line originate in gas surrounding one of the two
black holes.  In this context, the velocities of the two peaks require
that the mass ratio of the two black holes be of order unity.  One can
envisage alternative scenarios that are not constrained by the data we
have presented here.  For example, the nucleus may harbor a
supermassive binary in which one black hole is significantly more
massive than the other and accretes at a much higher rate.  In such a
scenario, an accretion disk around the more massive black hole can be
the source of the double-peaked line, but the location of the twin
peaks varies slightly as a result of the perturbation from the
companion.

\subsection {Jets and Outflows} 

Jets such as those proposed to produce double-peaked Balmer emission
lines must be quite different from those previously observed in AGNs.
If continuous flows, these jets have to be subrelativistic to produce
peaks at the correct wavelengths and have very broad opening angles to
produce substantial flux at zero velocity ($cf.$ the models of Zheng,
Veilleux, \& Grandi 1990), but still have sufficiently low velocity
gradients to produce substantial optical depth; this is a strong
constraint (Halpern $et~al.$ 1996).  Alternatively, the jet emission
might occur at the intersection of a rapid jet and an extremely
turbulent (with velocity width $\approx$ 1000 km s$^{-1}$), massive
cloud.  However, it is unclear why such a phenomenon might produce two
widely displaced peaks more commonly than one.

Although radial outflow models as a class are very difficult to
constrain, specific versions of such models may be tested
observationally.  In particular, Livio \& Pringle (1996) have pointed
out that an accretion disk, if present, will obscure the receding
parts of an outflow at small radii from its center.  This suggests
that if double-peaked lines are to be attributed to an outflow, they
must originate in parts of that outflow far enough from the center of
the disk not to be obscured.  Moreover, if a central source of
ionizing radiation powers the line emission from the outflow,
variations in the luminosity of the source will result in
corresponding changes in the flux of the line.  The difference in
light travel time between the two sides of the outflow and an observer
on the Earth dictates that the two sides of a double-peaked emission
line should not respond simultaneously, but rather with a delay of
order $0.7\; M_8$~yr.  In this scenario, with the assumption that the
flow accelerates along the jet, the profile variations should have a
very specific pattern: a change should appear at low velocities first
and then propagate towards high velocities.  This type of perturbation
should first appear on the blue side of the line profile, and then its
mirror image on the red side should follow with a delay as estimated
above.  In contrast, the seemingly orbiting source of emission we
observe in Arp 102B ($cf.~ \S 4.2$) first appears on the red side of
the line and moves from one side of the line to the other.  Any
identification of this variable emission with a phenomenon propagating
along both jets would demand considerable fine-tuning of time delays
between the two peaks and other physical conditions.  The variable
part of the emission not only passes from one side of the peak to the
other but also returns to its original peak at the correct time,
requiring some sort of repeated phenomenon in the jet with its period
closely matching the time delay between the peaks.

 \subsection {Spiral Shocks in a Disk}	 

Spiral shock models cannot easily accommodate the time scale of Arp
102B's periodic variation.  Though such models do predict variations
in the line profile that pass from one side of the rest wavelength to
the other, the period of such variations will be roughly the viscous
time scale.  For one-armed spiral waves (which the dispersion relation
strongly favors over higher order displacements), Kato (1983)
calculates a period
\begin{equation} 
\tau = {2 \pi \Omega_{K} r^{2} \over c_{s}^2 w_{*}}={6.6 \times 10^{4}
\xi^{1/2} \over T_{4} w_{*}} \quad {\rm yr}, 
\end{equation} 
where $r$ is the radius at which the time scale is defined,
$\Omega_{K}$ is the angular frequency of a Keplerian orbit at that
radius, $c_{s}$ is the sound speed, and $w_{*}$ is a parameter believed to
be less than one (equal to the viscosity parameter $\alpha$ in the
Shakura-Sunyaev model for a period of $2 \pi t_{visc}$).
Since $\xi$ is at least 3 for stable orbits about a compact body, this
is clearly much longer than the observed period of variation.
Furthermore, previously described spiral shock models generally
include a third, minor peak at some epochs, and the two major peaks
are expected to also move about in wavelength on the pattern time
scale; these do not fit the variations seen.

It is also conceivable that a corrugation wave, which causes a
periodic warping of the disk, might produce similar variations (Kato
1989).  Such a wave travels around the disk on roughly the
sound-crossing time,
\begin{equation} 
\tau_{cross} \approx {80 M_{8} \xi_{2} \over T_{4}^{1/2}} \quad {\rm yr},
\end{equation}
where $\xi_{2}$ is $\xi$ in units of $10^2$; this can match the
observed time scales only for masses around $10^{6} $M$_{\sun}$.  
A warp can also be produced in the disk by an irradiation-induced 
instability, described by Pringle (1996). The precession time scale of the 
warp is given by (following Storchi-Bergmann $et~al.$ 1997) 
\begin{equation} 
\tau_{prec}\approx 7\times 10^5\; \left({m_{disk}\over 10~{\rm M}_{\odot}}
\right)\; {M_8^{1/2}\over \xi_2^{1/2}\; L_{43}}~{\rm yr,}
\end{equation}
where $m_{disk}$ is the mass of the accretion disk (estimated to be 
of order 10~M$_{\odot}$) and $L_{43}$ is the X-ray luminosity in units
of $10^{43}$~erg~cm$^{-2}$~s$^{-1}$.  Both of the above time scales are too
long to account for the observed profile variability in Arp~102B.  Moreover, Pringle (1996) estimates that the radiation-driven instability would operate at radii of order 0.1 pc, much larger than the scales we are interested in here.

\section{Discussion and Conclusions} 

A model attributing the variability of the broad, double-peaked
H$\alpha$ profile of Arp 102B to a hot spot on a circular orbit is
consistent with several tests of the data over more than two complete
cycles in the red-to-blue flux ratio variations of the peaks.  Both
the modeling described here and other observations of Arp 102B (e.g.,
Halpern $et~al.$ 1996) agree with the hypothesis that the remaining
emission is due to a circular accretion disk.  It seems clear from the
discussion of the preceding section that the variable excess emission
is not associated with subrelativistic jets or spiral shock waves.
If such structures were indeed present, a separate origin for the
periodically varying part of the line would have to be invoked.  In
that case, the variable part of the line would still be emitted at a
radius within the range expected for the Balmer-line emitting part of
an accretion disk.  In such situations, however, our mass
determination would have an additional uncertainty, since the
inclination of the orbit of the source of the variable emission would
no longer be known.

If some materially coherent, orbiting source of emission did not lie
within a circular accretion disk, it is likely to have a substantially
elliptical orbit (as there are few strong circularizing effects
operating on time scales of a few years).  However, in such a case,
fitting the excess emission by assuming a hot spot along a circular
orbit should require substantially varying angular velocity ($i.e.$,
the slope of the angle vs. time curve), in contrast to what is
observed.  Thus, the assumption that the double peaks of Arp 102B are
produced by a circular accretion disk yields a natural explanation for
the constant angular velocity of a region of excess emission; this
provides further support for the accretion disk hypothesis.

The fact that the hot spot persists with roughly constant strength
through two orbits and then decays over a shorter time scale allows
some evaluation of the possible source of that excess emission.  If it
dissipates due to viscous processes, for instance, it must be
extremely compact compared to the disk as a whole, since the viscous
time is
\begin{equation}
\tau_{visc} = {l \over v_R} \approx {48.2 \over \xi^{1/2} M_8 \alpha
T_{4}}\; \left({l \over R_{\sun}}\right) \quad {\rm yr},
\end{equation}
where $l$ is the characteristic radial size of the disturbance and
$v_R$ is the radial velocity of gas within the disk relative to its
center.

Disturbances can propagate azimuthally (through 360$^\circ$) or
vertically (through one scale height) on the Keplerian orbit period;
from the behavior of $R$, we may conclude that this is roughly 2.2
years, somewhat longer than the period over which the damping occurs.
The hot spot may dissipate thermally over the time scale
\begin{equation} 
\tau_{th} \approx {\theta \over \Omega_{K} \alpha} \; \approx {0.1 M_8
\xi_2^{3/2} \theta \over \alpha} \; \quad {\rm yr},
\end{equation}
where $\theta$ is the azimuthal angular extent of the hot spot.  The
hot spot appears to decrease in intensity over an appreciable fraction
of a year; thus, for this mechanism to operate, it should have
substantial azimuthal extent or $\alpha$ should be much less than
unity.  Heat waves will propagate in the time it takes sound to travel
along the spot,
\begin{equation}
\tau_{sound} \approx {80 M_{8} \xi_{2} \theta \over T_{4}^{1/2}} \;
\quad {\rm yr},
\end{equation} 
requiring a very small angular extent (which, based upon the values of
$\sigma$ providing best fits, is almost certainly less than
$20^{\circ}$) or a high internal temperature for the hot spot to
dissipate its energy.

One constraint on the radial structure of the hot spot may be found
from its azimuthal evolution, since we expect a large spot to be
smeared out by Keplerian shear.  In a time $\Delta t$ the azimuthal
extent of a hot spot at radius $r$ and radial extent $\Delta r$ 
will increase by 
\begin{equation} 
{\Delta\theta\over 2\pi} = {3\over 2}\; {\Delta t \over P}\
{\Delta r\over r},
\end{equation} 
where $P$ is the local Keplerian period.  Because the hot spot we have
considered in our model does not appear to evolve azimuthally over the
course of two revolutions (the average value of $\sigma$ determined
for the second cycle of the hot spot, $8^{\circ}$, is no greater than
that for the first), it must be very compact ($\Delta r \ll r$).  This
can be the case if, for example, the hot spot resulted from the
oblique impact of a star on the disk, which created a trail of
substantial azimuthal extent but small radial extent (note that
$r\approx 455~r_g \approx 10^5\; M_8\; {\rm R}_{\odot}$).
Alternatively, this may indicate that $\sigma$ does not measure the
actual angular extent of the hot spot, which would presumably be
achieved by such a velocity shear, but instead its velocity structure.

The most likely scenarios for a hot spot are those in which there is
excess material or energy concentrated in some location with,
possibly, some extent in azimuthal angle but little in radius.  The
most common potential sources of such changes present in an AGN
environment are stars (Blandford \& Rees 1992), but the exact
mechanism is unclear.  Stars passing through the disk are expected to
remove material, rather than to deposit any, and the material
removed should quickly dissipate after it is no longer shielded from
the hot, luminous central parts of the disk by an optically thick
medium (Zurek, Siemiginowska, \& Colgate 1994).  Stars should be
brought into orbits lying within the disk through the momentum loss
that occurs in passages through it; however, due to the concentration
of material within the disk, such stars should not then leave its
central plane.  Therefore, scenarios in which the strong variations in
$R$ occur while a star orbits within the disk and subsequently
decrease in strength when it begins to leave are untenable.

An alternative scenario is one in which the excess emission comes from
a sizable vortex orbiting within the disk (Abramowicz $et~al.$ 1992).
Such a vortex may possess a substantial velocity gradient, explaining
the apparent width of the hot spot.  However, examination of the power
spectrum of the complete set of $R$ values reveals only a single
strong peak (with a period corresponding to that measured above) and
its aliases; models in which there exists a cascade of large numbers
of vortices at many scales predict a power spectrum of roughly
power-law form (Abramowicz $et~al.$. 1991).

It is clear that some type of temporary periodic phenomenon occurs,
whatever its origin, and that the most likely way for it to cycle
between being blueshifted and redshifted is if it follows an orbit
around the central body.  The self-consistency of a model in which the
hot spot follows a circular orbit is encouraging.  The constant
angular velocity of the hot spot supports the hypothesis that it is
within a circular accretion disk, as independent elliptical orbits do
not circularize quickly compared to the orbital period.  Thus, a model
in which the two peaks of Arp 102B arise in a circular accretion disk
seems to be the simplest that can explain the observed phenomena.
Even if this model is incorrect, the consistency of the behavior of
the variable part of the emission line with a circular orbit suggests
that our mass determination should be correct up to an uncertainty in
the inclination.

Our measurement of the mass of a possible black hole at the center of
Arp 102B is subject to a straightforward test.  Unless the hot spot
observed was a particularly rare event, other such phenomena should
appear in the future and may be subjected to similar analysis.  If the
spectrum if Arp 102B is observed sufficiently often, an independent
mass measurement should result.  If our model is correct, those
measurements should be consistent, even though a new hot spot may be
located at a different radius and have different intensity than the
one we observed.

Contrary to previous claims ($e.g.$, Gaskell 1996; Miller \& Peterson
1990), the variability of the line profile does not necessarily
invalidate the accretion disk hypothesis.  Instead, it may provide
additional information about the structure and properties of the disk.
In the particular case of Arp 102B, the pattern of variability leads
to a dynamical measurement of the mass of the central black hole.
This method may be applicable to other double-peaked emitters; several
have exhibited substantial variation in the blue-to-red flux ratio,
for instance, and the variations of their line profiles may not be
fitted in detail by the previously posited elliptical accretion disk
models.  With well-sampled monitoring such as that conducted for 3C
390.3 (Veilleux \& Zheng 1991), more mass measurements of this quality
might be made.

\acknowledgments

Most of the data presented in this paper were collected at Kitt Peak
National Observatory and Lick Observatory.  We are grateful to the
members of the staffs for their expert assistance in carrying out the
observations, as well as to numerous colleagues who have contributed their efforts to this project.  M. E. and J. A. N. acknowledge Hubble Fellowship grant
HF-01068.01-94A and a National Science Foundation Fellowship,
respectively.  This work was also supported by NASA grant G0-06097-94A
from the Space Telescope Science Institute (operated by AURA, Inc.,
under NASA contract NAS5-26555).

\clearpage

\clearpage

\begin{deluxetable}{llrrr}
\tablewidth{31pc}
\tablecaption{Journal of Observations \tablenotemark{a}}
\tablehead{
\tablevspace {-0.5em} \multicolumn{1}{c}{UT Date} & \multicolumn{1}{c}{Telescope} & \multicolumn{2}{c}{Exposure Time (s)} & \nl
\tablevspace {-0.5em}
\multicolumn{1}{c}{} & \multicolumn{1}{c}{} & \multicolumn{2}{l}{\hrulefill} & \nl
\tablevspace {-0.25em} \multicolumn{1}{c}{} & \multicolumn{1}{c}{} & \multicolumn{1}{c}{H${\alpha}$} & \multicolumn{1}{c}{H${\beta}$} & \nl
\tablevspace {-1em} \multicolumn{4}{c}{}
}
\startdata
1983 Jun  6 & Palomar 5 m & \tablenotemark{b} & \tablenotemark{b} & \nl 
1983 Sep 15 & Palomar 5 m & \tablenotemark{b} & \tablenotemark{b} & \nl 
1985 Jun 28 & Palomar 5 m & 2500 & 2500  & \nl
1986 Jul 13 & Lick 3 m   & 3000  & 3000  & \nl
1987 May  4 & Lick 3 m   & 1000  & \dots & \nl
1987 Aug  8 & Lick 3 m   & 1300  & \dots & \nl
1987 Aug 12 & Lick 3 m   & 1600  & \dots & \nl 
1989 Apr 27 & Lick 3 m   &  900  & \dots & \nl
1989 Jul  1 & KPNO 2.1 m & 3600  & \dots & \nl
1989 Jul  3 & KPNO 2.1 m & \dots & 1800  & \nl 
1989 Jul  5 & KPNO 2.1 m &  835  &  835  & \nl
1989 Jul 10 & Lick 3 m   &  900  & \dots & \nl
1989 Nov  4 & MDM 2.4 m  & 1800  & \dots & \nl
1989 Nov  7 & MDM 2.4 m  &  435  & \dots & \nl
1989 Dec  1 & Lick 3 m   & 1600  & 1600  & \nl
1990 Feb 23 & KPNO 2.1 m & 5400  & \dots & \nl
1990 Feb 24 & KPNO 2.1 m & \dots & 5400  & \nl
1990 May 30 & KPNO 2.1 m & \dots & 2700  & \nl
1990 Jul 17 & Lick 3 m   & 1800  & \dots & \nl
1990 Aug 30 & Lick 3 m   & 1800  & \dots & \nl
1990 Nov 11 & Lick 3 m   & 1800  & \dots & \nl
1991 Jun 17 & KPNO 2.1 m & 9790  & \dots & \nl
1991 Jun 18 & KPNO 2.1 m & 1897  & \dots & \nl
1991 Jun 19 & KPNO 2.1 m & 7200  & \dots & \nl
1991 Jun 20 & KPNO 2.1 m & 7200  & \dots & \nl
1991 Jul  3 & KPNO 2.1 m & 7200  & \dots & \nl
1991 Jul  4 & KPNO 2.1 m & 4950  & \dots & \nl
1991 Aug  5 & Lick 3 m   & 1500  & \dots & \nl
1991 Oct 31 & Lick 3 m   & 1800  & \dots & \nl
1992 Apr 21 & Lick 3 m   & 1200  & \dots & \nl
1992 May  6 & KPNO 2.1 m & 7200  & 7200  & \nl
1992 May 13 & KPNO 2.1 m & \dots & 4800  & \nl 
1992 May 14 & KPNO 2.1 m & \dots & 4800  & \nl 
1992 May 16 & KPNO 2.1 m & 3600  & 3600  & \nl 
1992 Aug  3 & Lick 3 m   & \dots & 1800  & \nl  
1992 Oct  3 & Lick 3 m   & 1800  & 1800  & \nl
1992 Nov 19 & Lick 3 m   & 1200  & 1200  & \nl
1993 Apr 14 & Lick 3 m   &  900  &  900  & \nl
1993 May 16 & KPNO 2.1 m & 3600  & 3600  & \nl
1993 Jun 28 & Lick 3 m   &  900  &  900  & \nl
1993 Jul 28 & Lick 3 m   & 2100  & 2100  & \nl
1993 Sep 10 & Lick 3 m   & 3000  & 3000  & \nl
1993 Sep 25 & Lick 3 m   & 1800  & 1800  & \nl
1993 Oct 22 & Lick 3 m   & 1800  & 1800  & \nl
1993 Nov  8 & Lick 3 m   & 3000  & 3000  & \nl
1994 Apr 18 & Lick 3 m   & 1200  & 1200  & \nl
1994 Jun 16 & Lick 3 m   & 2700  & 2700  & \nl
1994 Jul  4 & KPNO 2.1 m & 3600  & \dots & \nl 
1994 Jul  5 & KPNO 2.1 m & 3600  & \dots & \nl 
1994 Jul 15 & Lick 3 m   & 3000  & 3000  & \nl
1994 Aug  4 & Lick 3 m   & 3600  & 3600  & \nl
1994 Sep  3 & Lick 3 m   & 3600  & 3600  & \nl
1994 Oct  1 & Lick 3 m   & 4800  & 6600  & \nl
1994 Nov 12 & Lick 3 m   & 3600  & 2800  & \nl
1995 Jan 23 & KPNO 2.1 m & 1800  & \dots & \nl
1995 Mar 25 & Lick 3 m   & 1500  & 1500  & \nl
1995 Jun  3 & KPNO 2.1 m & 3600  & \dots & \nl 
1995 Jun  4 & KPNO 2.1 m & 3600  & 3600  & \nl
1995 Sep 26 & Lick 3 m   & 1800  & 1800  & \nl
1996 Feb 13 & KPNO 2.1 m & 3200  & 3200  & \nl
1996 Jun 13 & KPNO 2.1 m & 1711  & 1711  & \nl

\tablenotetext{a}
{The UV Schmidt spectrograph (Miller \& Stone 1987)
was used for all Lick Observatory observations listed here until the
end of 1991; the Kast spectrograph (Miller \& Stone 1993) was used
thereafter.  At Palomar Observatory, we used the Double Spectrograph
(Oke \& Gunn 1982).}  
\tablenotetext{b}{The 1983 Palomar 5 m spectra
are composites of several exposures with effective exposure
times that are different in different spectral regions} 
\enddata
\end{deluxetable}

\clearpage

\begin{deluxetable}{ccc}
\tablewidth{31pc}
\tablecaption{Rest Wavelength of Blue Peak}
\tablehead{
\tablevspace {-0.5em}
\multicolumn{1}{c}{Year} & \multicolumn{1}{c}{Pogson's Method} & 
\multicolumn{1}{c}{Gaussian Method}  \nl
\multicolumn{1}{c}{} & \multicolumn{1}{c}{\AA} & \multicolumn{1}{c}{\AA} \nl
\tablevspace {-0.5em} } \startdata 1982.43 & 6451\tablenotemark{a} &
\dots \nl 1985.49 & 6460.99 & 6459.02 \nl 1986.53 & 6445.02 & 6452.96
\nl 1987.34 & 6456.28 & 6458.06 \nl 1987.61 & 6453.17 & 6455.81 \nl
1989.32 & 6451.14 & 6452.61 \nl 1989.45 & 6445.29 & 6450.93 \nl
1989.50 & 6445.28 & 6450.93 \nl 1989.51 & 6442.61 & 6449.56 \nl
1989.52 & 6447.27 & 6450.58 \nl 1989.84 & 6448.12 & 6449.97 \nl
1989.85 & 6453.88 & 6450.37 \nl 1989.92 & 6444.70 & 6448.28 \nl
1990.15 & 6442.09 & 6449.17 \nl 1990.41 & 6449.46 & 6453.35 \nl
1990.54 & 6448.38 & 6454.19 \nl 1990.66 & 6444.62 & 6454.06 \nl
1990.86 & 6462.76 & 6459.21 \nl 1991.46 & 6439.54 & 6446.19 \nl
1991.46 & 6443.08 & 6446.01 \nl 1991.46 & 6441.50 & 6445.80 \nl
1991.47 & 6441.89 & 6445.86 \nl 1991.50 & 6441.03 & 6445.77 \nl
1991.50 & 6439.82 & 6445.07 \nl 1991.59 & 6443.26 & 6445.31 \nl
1991.83 & 6443.50 & 6448.06 \nl 1992.30 & 6455.08 & 6453.72 \nl
1992.35 & 6447.96 & 6450.49 \nl 1992.76 & 6447.99 & 6450.99 \nl
1992.88 & 6448.53 & 6452.70 \nl 1993.28 & 6442.88 & 6454.19 \nl
1993.37 & 6450.10 & 6454.42 \nl 1993.49 & 6450.01 & 6454.43 \nl
1993.57 & 6447.22 & 6452.24 \nl 1993.69 & 6451.62 & 6451.42 \nl
1993.73 & 6450.13 & 6452.09 \nl 1993.81 & 6450.20 & 6450.74 \nl
1993.85 & 6444.53 & 6449.09 \nl 1994.29 & 6447.15 & 6452.81 \nl
1994.46 & 6444.48 & 6450.06 \nl 1994.51 & 6443.25 & 6448.78 \nl
1994.51 & 6444.26 & 6449.56 \nl 1994.54 & 6444.32 & 6449.47 \nl
1994.59 & 6447.08 & 6450.97 \nl 1994.67 & 6447.89 & 6450.77 \nl
1994.69 & 6447.87 & 6450.77 \nl 1994.75 & 6447.65 & 6450.94 \nl
1994.86 & 6445.84 & 6451.48 \nl 1995.06 & 6443.90 & 6451.27 \nl
1995.23 & 6448.42 & 6452.15 \nl 1995.23 & 6446.65 & 6453.84 \nl
1995.42 & 6451.27 & 6452.71 \nl 1995.42 & 6452.25 & 6452.58 \nl
1995.73 & 6454.60 & 6455.81 \nl 1996.12 & 6457.74 & 6457.55 \nl
1996.45 & 6451.57 & 6453.34 \nl 
\enddata 
\tablenotetext{a}
{Measurement was made by eye from the spectrum presented by Stauffer $et~al.$
(1983).}
\end{deluxetable}

\clearpage

\begin{deluxetable}{lrrrrrr}
\tablewidth{31pc} \tablecaption{Fitted Model Parameters} \tablehead{
\multicolumn{1}{c}{UT Date} & \multicolumn{1}{c}{$\theta$} &
\multicolumn{1}{c}{$I$} & \multicolumn{1}{c}{$\sigma$} &
\multicolumn{1}{c}{$\xi_i$} & \multicolumn{1}{c}{$\xi_o$} &
\multicolumn{1}{c}{$L_{spot} / L_{disk}$} \nl
}
\startdata
1990 Jul 17 & \dots &   0 & \dots & 305 &  730 & \dots \nl
1990 Nov 11 & 115 & 150 &  3.0 & 295 &  940 & 0.0065 \nl
1991 Jun 18 & 270 & 180 &  8.0 & 280 &  940 &  0.019 \nl
1991 Jul 04 & 270 & 170 &  8.0 & 275 &  940 &  0.018 \nl
1991 Aug 05 & 270 & 150 &  4.0 & 305 &  790 &  0.010 \nl
1991 Oct 31 & 230 & 150 &  19. & 240 & 1000 &  0.030 \nl
1992 May 06 &  30 & 190 &  8.0 & 280 &  750 &  0.023 \nl
1992 Oct 03 &  20 & 170 &  5.0 & 305 &  770 &  0.014 \nl
1992 Nov 19 &  28 & 235 &  5.0 & 285 &  820 &  0.017 \nl
1993 Apr 14 & 225 & 140 &  8.0 & 320 &  880 &  0.019 \nl
1993 May 16 & 185 & 205 &  10. & 295 &  820 &  0.032 \nl
1993 Jun 28 & 196 & 235 &  10. & 305 &  900 &  0.037 \nl
1993 Jul 28 & 220 & 145 &  8.0 & 310 &  880 &  0.019 \nl
1993 Sep 25 & 295 & 160 &  7.0 & 320 &  880 &  0.019 \nl
1993 Oct 22 & 260 & 105 &  9.0 & 320 &  880 &  0.016 \nl
1994 Apr 18 & 175 & 150 &  15. & 345 &  650 &  0.056 \nl
1994 Jul 05 & 140 &  90 &  2.5 & 400 &  500 &  0.015 \nl
1994 Sep 03 &  20 & 115 &  7.0 & 350 &  630 &  0.021 \nl
1994 Oct 01 &  20 & 150 &  5.0 & 335 &  700 &  0.016 \nl
1994 Nov 12 &  45 &  65 &  5.0 & 350 &  600 & 0.0092 \nl
1995 Jan 23 &  65 &  35 &  5.0 & 350 &  650 & 0.0045 \nl
1995 Mar 25 &  25 &  70 &  10. & 325 &  750 &  0.014 \nl
1995 Jun 03 & 270 &  20 &  10. & 315 &  790 & 0.0035 \nl
1996 Jun 13 & 240 &  60 &  4.0 & 305 &  790 & 0.0040 \nl
\enddata
\end{deluxetable}

\newpage
\begin{figure}
\epsfxsize=6truein
\epsfbox[0 0 600 300]{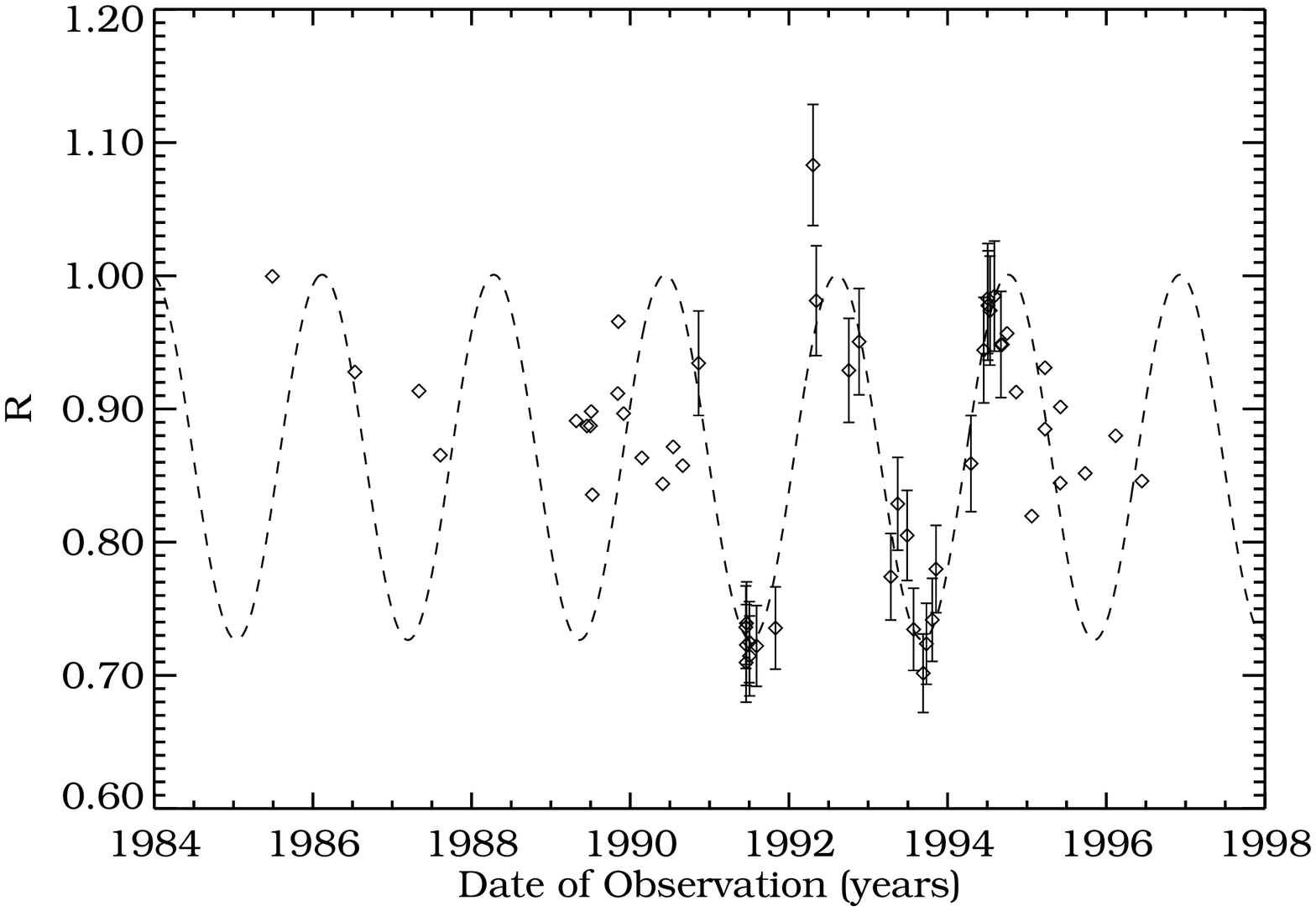}
\vfill
\noindent Figure 1 - The evolution of the normalized red-to-blue flux
ratio, $R$, in spectra of Arp 102B.  Beginning around November 1990,
$R$ varies sinusoidally for several years before returning to its
value before the variation began.  The sinusoidal curve is a
least-squares best fit to those data having error bars shown
(corresponding to the epochs from 1990 November 11 through 1994
September 3).  Comparison of the residuals for data at similar phases
indicates that there is no obvious substantial change in the period of
variation during this time.
\end{figure}

\newpage
\begin{figure}
\epsfysize=8truein
\epsfbox{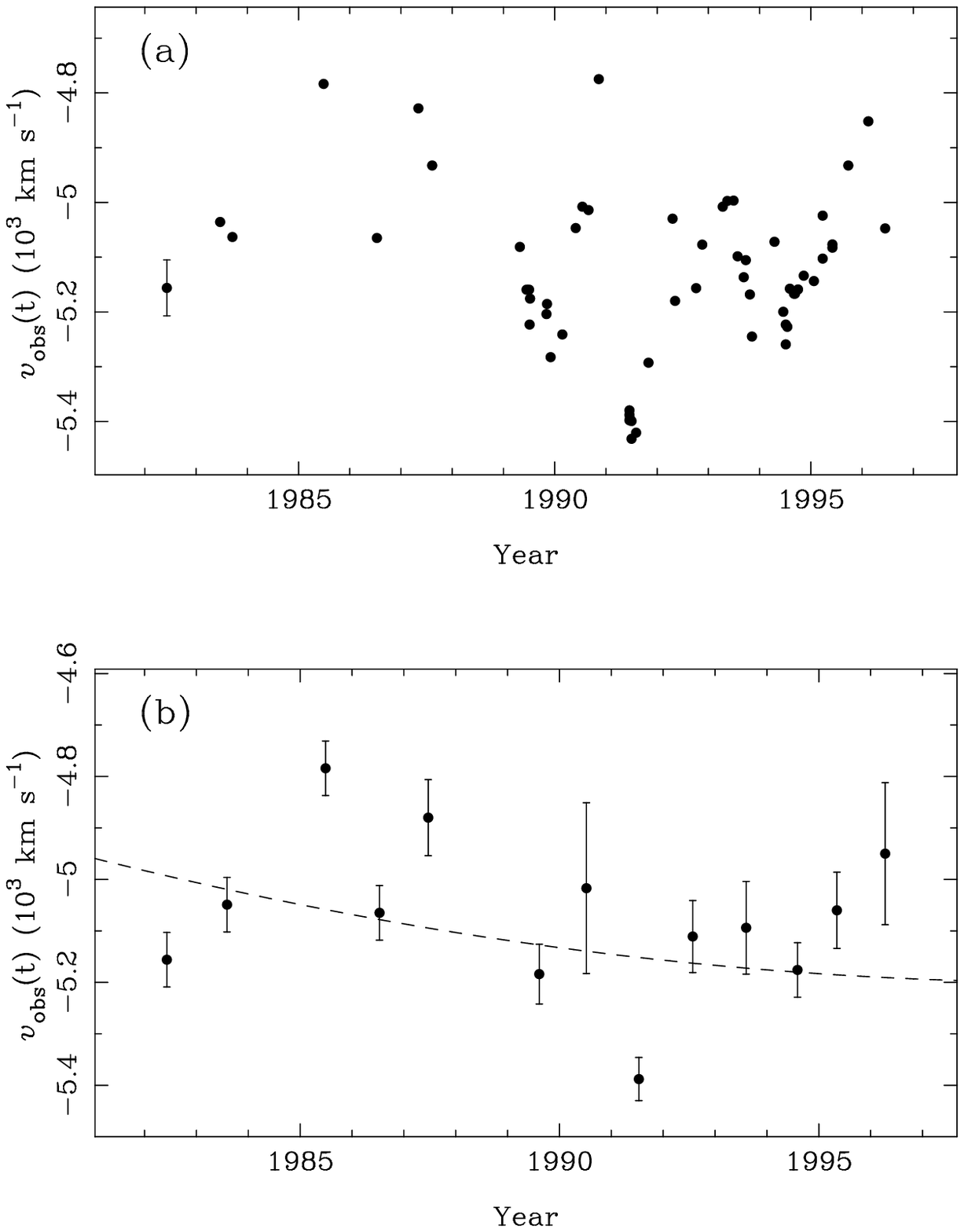}
\vfill
\noindent Figure 2. -- (a) The variation of the peak velocity of the
blue peak as measured from individual spectra using the Gaussian
fitting method.  A typical error bar of 100~km~s$^{-1}$ is shown on
the very first point in the sequence for reference.  (b) The annually
averaged blue peak velocities.  The error bars correspond to the
dispersion of the velocities measured from individual spectra during
that year.  The dashed line is the best-fitting sinusoid of equation 4
through these points, which has a period of 390 years and an amplitude
of 5200~km~s$^{-1}$.
\end{figure}

\newpage
\begin{figure}
\epsfxsize=6truein
\epsfbox[0 0 600 300]{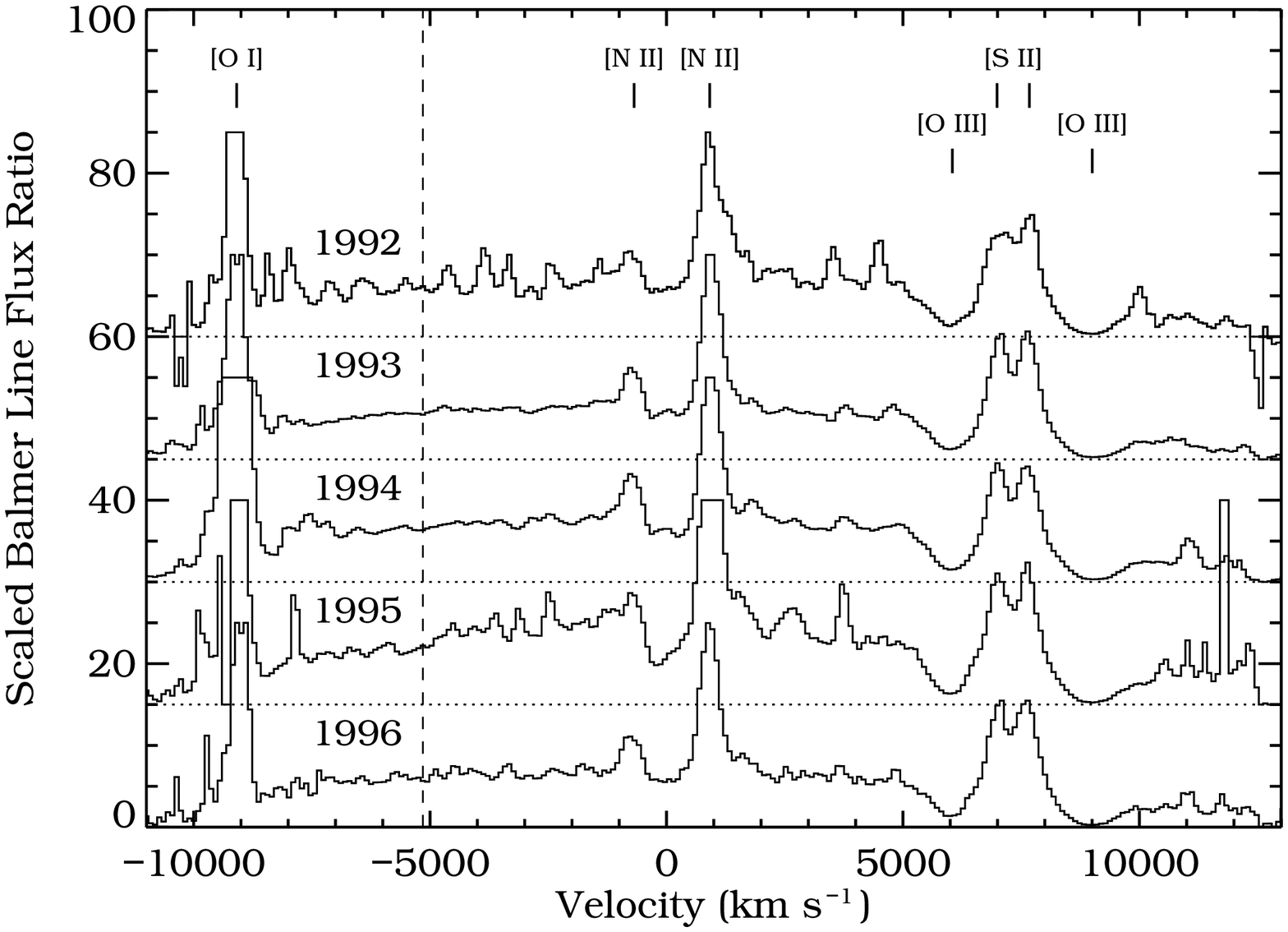}
\vfill
\noindent Figure 3 - The rescaled ratios of the annual average of the
H$\alpha$ flux for spectra in a given year to the annual average of
the H$\beta$ flux in that year.  Each average spectrum is indicated by
a solid line and is offset by 15 units from spectra of the previous
and following years; the zero point for each spectrum is indicated by a
dotted line.  The dashed line indicates the mean velocity of the blue
peak in the H$\alpha$ spectra; all excursions in peak velocity from
this value were less than 500 km s$^{-1}$.  The strongest peaks and
valleys are associated with the indicated narrow lines near either
H$\alpha$ and H$\beta$.
\end{figure}

\newpage
\begin{figure}
\epsfysize=6truein
\epsfbox{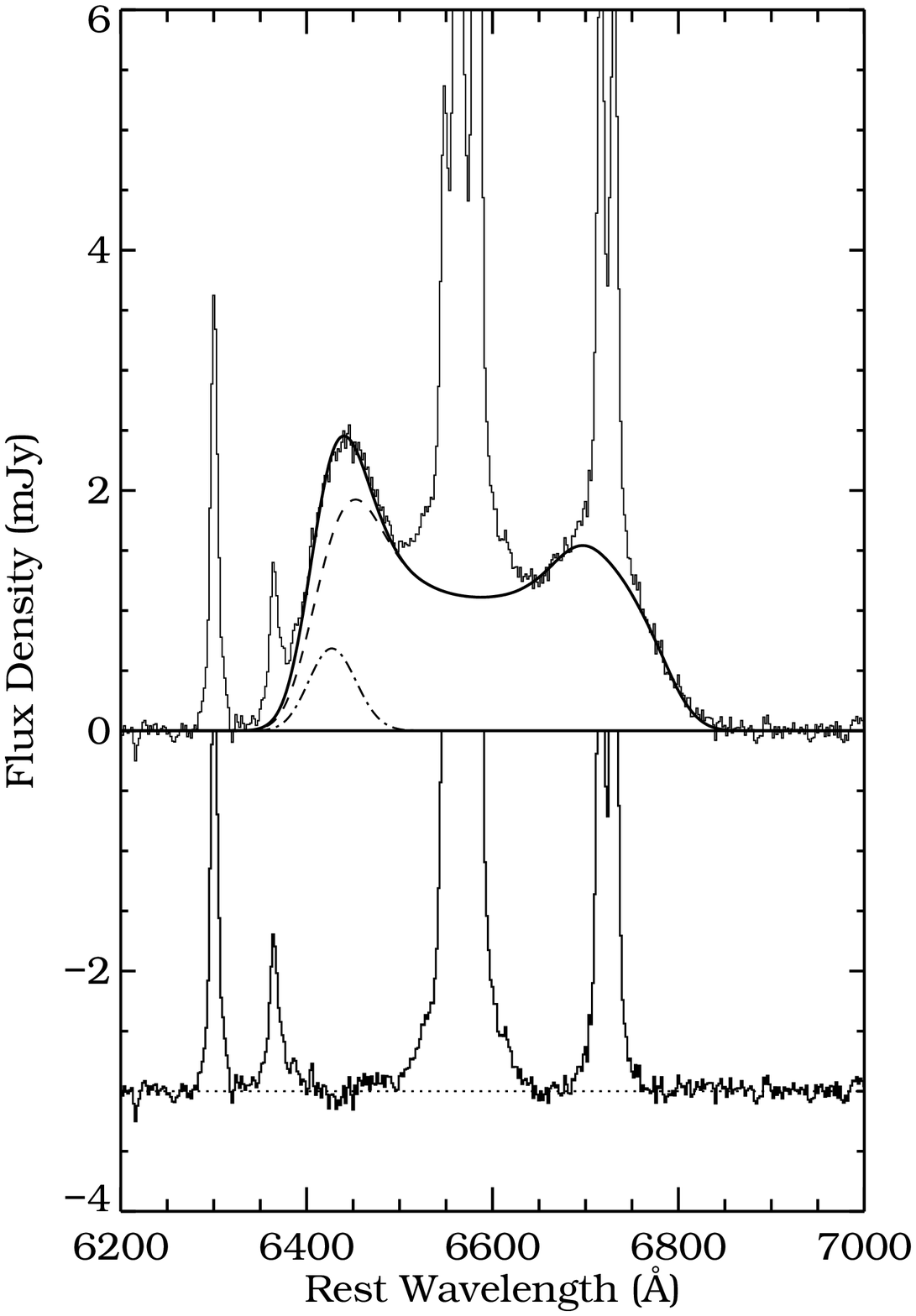}
\vfill
\noindent Figure 4 - An example of a fit to a spectrum using our
model.  The actual spectrum (taken 1991 June 17) and our model fit are
indicated with solid curves (some of the narrow emission lines have
been truncated in flux density).  The dashed line is the flux density
in our model fit due to the circular, axisymmetric accretion disk; the
dot-dashed line indicates the flux emitted by the hot spot.  Finally,
the thick solid line indicates the residual when our model fit is
subtracted from the actual spectrum, offset by -3 mJy in flux density.
Note the resemblance of this residual to the spectrum of a typical
broad-line radio galaxy.  The model parameters of the fit are
$\xi_i$=280, $\xi_o$=940, $\xi_{hot}$=455, $\theta=270^{\circ}$,
$I$=180, and $\sigma=8^{\circ}$.
\end{figure}

\newpage
\begin{figure}
\epsfysize=6truein
\epsfbox{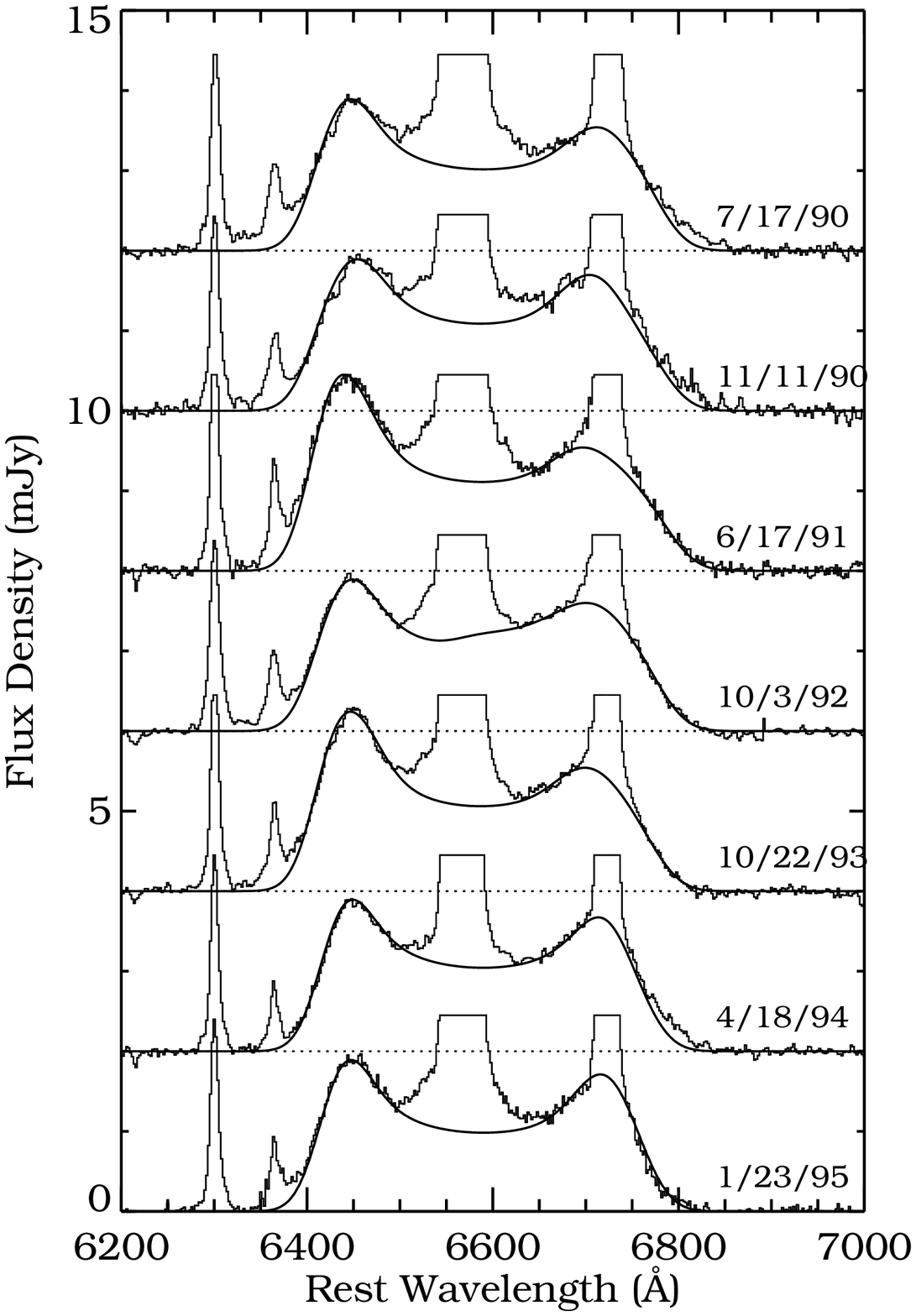}
\vfill
\noindent Figure 5 - A set of spectra representative of the evolution
of Arp 102B.  Spectra are separated by arbitrarily adding a multiple of 2
mJy; the zero point for each spectrum is indicated by a dotted line.  The
first spectrum, from 1990 July 17, includes no hot spot component.
However, the spectra from 1990 November 11 through 1995 January 23
include a hot spot at an azimuth angle of 115, 270, 160, 280, 5, and
115 degrees, respectively, corresponding to hot spots centered at
roughly 6842, 6255, 6668, 6260, 6590, and 6842 \AA.  Substantially
different spectra are all fitted by the model.  The amplitude of the
hot spot had begun to decrease by 1995.
\end{figure}

\newpage
\begin{figure}
\epsfxsize=6truein
\epsfbox[0 0 600 300]{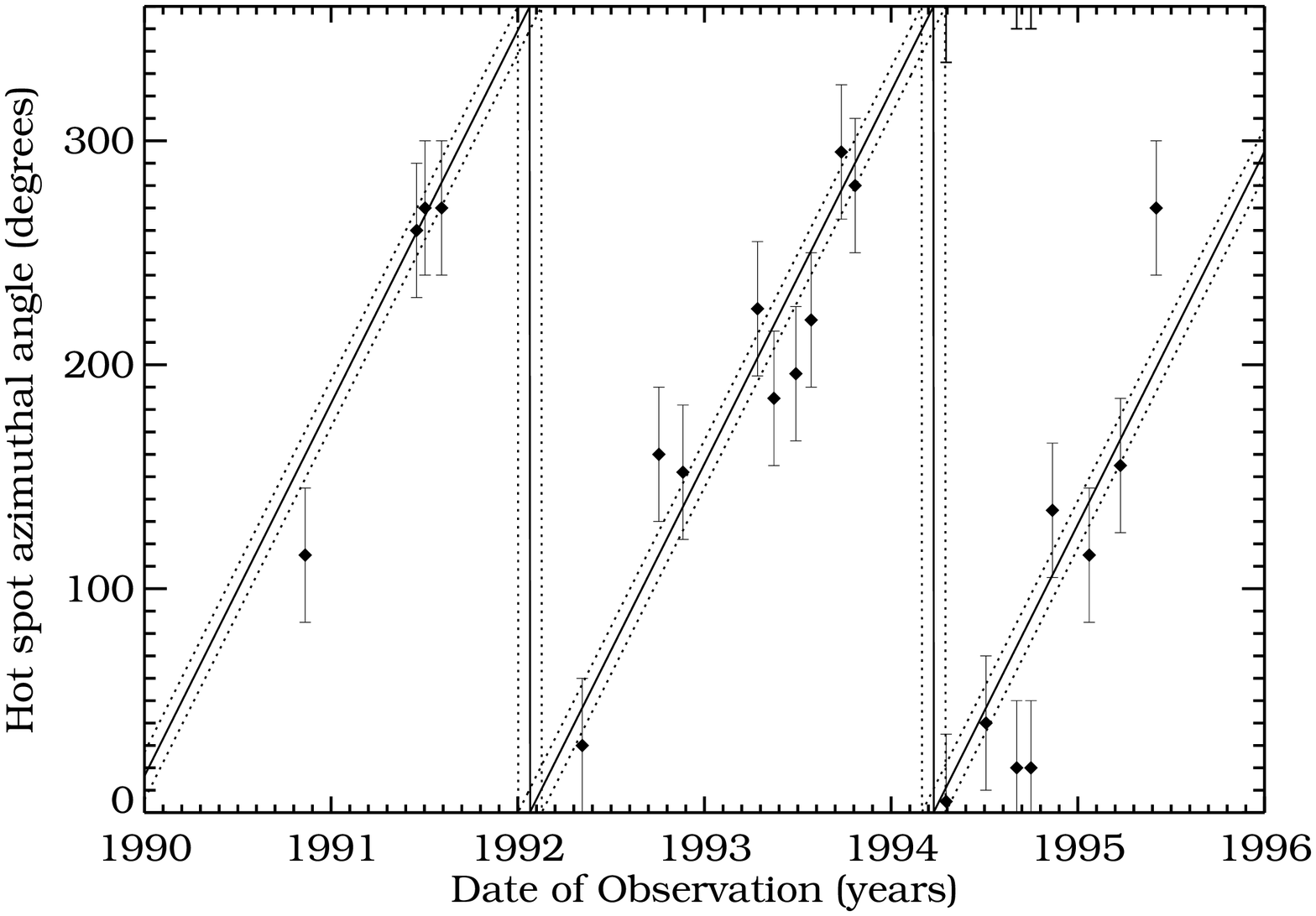}
\vfill
\noindent Figure 6 - The azimuthal angle of the hot spot measured from
our model fits.  The solid line is not a fit to the data shown here,
but instead is the increasing phase angle of the sinusoid fit shown in
Figure 1; the dotted lines indicate the 1 $\sigma$ error bounds of
that fit.  At each epoch, there are two possible values of the
azimuthal angle due to degeneracies in our model; the possible angle
which lies closest to the line is plotted here.  Note that there is no
systematic pattern in the scatter of data about the line from
different epochs with similar phase.  It is thus unlikely that the
angular velocity varies substantially but periodically (as would be
expected for an object on an elliptical orbit).
\end{figure}

\end{document}